\documentclass{article}

\usepackage{meta}
\usepackage{amssymb}
\usepackage{amsmath}
\usepackage{epsfig}

\title{SQUID Metamaterials: Tuneability and Multistability} 
\makeatletter
\def\name#1{\gdef\@name{#1\\}}
\makeatother
\name{{\bf \large  G. P. Tsironis$^{1,2}$, N. Lazarides$^{1,2}$ }}
\address{
$^1$Department of Physics, University of Crete, P.O. Box 2208,
71003 Heraklion, Greece \\
$^2$Institute of Electronic Structure and Laser,
Foundation for Research and Technology-Hellas, \\
P.O. Box 1527,  71110 Heraklion, Greece \\
*corresponding author, E-mail: {\tt nl@physics.uoc.gr}
}

\begin{document}
\maketitle

\begin{abstract}
An overview of several dynamic properties of SQUID metamaterials is given in 
the presence of both constant and alternating magnetic field. The total current 
as a function of the driving frequency exhibits hysteretic effects which are 
favored by low levels of disorder. Multistability in the current states leads 
to multiple magnetic responses with different value of magnetic permeability.
SQUID metamaterials exhibit wide-band tuneability which is periodic with the 
applied constant magnetic field; the numerical calculations reproduce fairly
well recent experimental results. Current work also reveals the possibility
for wave transmission through nonlinear bands, which is briefly discussed.  
\end{abstract}

%
\section{Introduction}
Superconducting metamaterials \cite{Anlage2011}, a particular class of artificial 
media that rely on the extraordinary properties acquired by superconductors
at sufficiently low temperatures, have been recently attracted great attention
(see e.g., reference \cite{SUST-Focus}).
In contrast to conventional metamaterials, which suffer from high losses due to
energy dissipation in their metallic parts, superconducting metamaterials exhibit
low Ohmic losses and, moreover, intrinsic nonlinearities due to the extreme 
sensitivity of the superconducting state to external fields \cite{Zheludev2010}. 
Nonlinearity provides tuneability of the effective electromagnetic properties 
of superconducting metamaterials \cite{Chen2010}, while the high degree of 
controllability through external fields gives them remarkable   switching 
capabilities \cite{Gu2010}.

The direct superconducting analogue of a nonlinear split-ring resonator is the 
rf SQUID 
(rf Superconducting QUantum Interference Device) that consists of a superconducting
ring interrupted by a Josephson junction, shown schematically in Fig. 1(a) along
with its equivalent circuit [Fig. 1(b)].
A planar periodic array of rf SQUIDs operate as 
a strongly nonlinear and tuneable metamaterial \cite{Du2006,Lazarides2007}, 
with unusual magnetic properties \cite{Lazarides2008a,Lazarides2013b} and reduced
losses. The feasibility of using rf SQUIDs as basic 
elements for the construction of superconducting thin-film metamaterials
\cite{Jung2013,Ghamsari2013}, and the tunability of SQUID metamaterials over
a broad frequency band \cite{Butz2013a,Trepanier2013} has been recently demonstrated.

In the present work, an overview of several theoretical results on the complex 
dynamics of SQUID metamaterials, related to multistability and self-organization,
wide-band tuneability, and nonlinear wave transmission, is given. 
The dynamics of a SQUID metamaterial is considered in the weak coupling
approximation, while single SQUIDs are described by the equivalent circuit of 
Fig. 1(b); in that picture, the Josephson element in each SQUID is regarded as
a resistively and capacitively shunted junction.

\begin{figure}[h!]
\centerline{\epsfig{figure=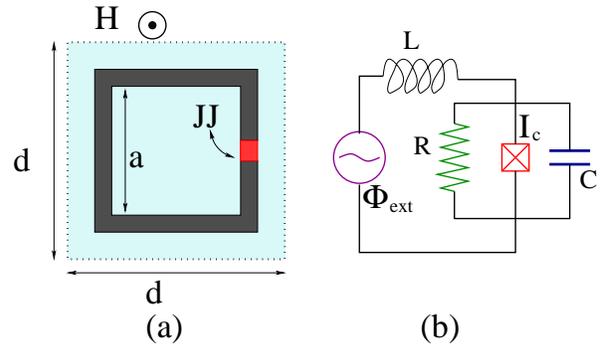,width=80mm}}
\caption{(a) Schematic drawing of an rf SQUID in an alternating magnetic field.
Equivalent electrical circuit for the rf SQUID shown in (a)}
\label{fig:1}
\end{figure}
\section{SQUID Metamaterial Equations}
Consider a two-dimensional tetragonal lattice of identical SQUIDs in a spatially 
uniform magnetic field that may have both constant (dc) and alternating (ac) 
components, ${\bf H} (t)$. In the following we adopt the description (and notation) 
in Ref. \cite{Lazarides2013b}, thus summarizing the essential building blocks of 
the model in a self-contained manner, yet omitting unnecessary details.
The dynamic equations for the fluxes in the $(n,m)$ SQUID ring, $\phi_{n,m}$, in the weak
coupling approximation, are given by \cite{Lazarides2008a,Lazarides2013b}
\begin{eqnarray}
\label{2.01}
  \ddot{\phi}_{n,m} &+&\gamma \dot{\phi}_{n,m} +\phi_{n,m} +\beta\, \sin( 2 \pi \phi_{n,m} )
 \nonumber \\
  &-&\lambda_x ( \phi_{n-1,m} +\phi_{n+1,m} ) 
 \nonumber \\
  &-&\lambda_y ( \phi_{n,m-1} +\phi_{n,m+1} )
   = \phi_{eff} , 
\end{eqnarray}
where the overdots denote differentiation with respect to the normalized time 
$\tau$, $\lambda_x$ ($\lambda_y$) are the coupling coefficients in the $x$ 
($y$) direction, $\beta ={L I_c}/{\Phi_0} =\beta_L /2\pi$ 
is the rescaled SQUID parameter, and $\gamma$ a dimensionless loss coefficient. 
The frequency and time in Eq. (\ref{2.01})
have been normalized to the corresponding inductive-capacitive SQUID
frequency $\omega_0 =1/\sqrt{LC}$ and its inverse $\omega_0^{-1}$, respectively,
while fluxes are normalized to the flux quantum, $\Phi_0$.
The effective applied flux $\phi_{eff}$ is expressed through the external flux,
$\phi_{ext} = \phi_{dc}  +\phi_{ac} \cos(\Omega t )$, as
$\phi_{eff} =[1-2(\lambda_x +\lambda_y)] \phi_{ext}$, 
where $\Omega$ is the driving frequency. The current flowing in the $(n,m)$ 
SQUID, $i_{n,m}$, normalized to the critical current of the junction, $I_c$,
is given in the same approximation as in equation (\ref{2.01}), as
\begin{eqnarray}
\label{2.03}
   i_{n,m} =\frac{1}{\beta} \left\{ \phi_{n,m} - \phi_{eff} 
           -\lambda_x ( \phi_{n-1,m} + \phi_{n+1,m} ) \right. 
\nonumber \\ \left.
           -\lambda_y ( \phi_{n,m-1} + \phi_{n,m+1} ) \right\}.
\end{eqnarray}
The dispersion of low-amplitude flux waves in SQUID metamaterial is given by
\begin{eqnarray}
  \label{2.04}
   \Omega \equiv \Omega_{\bf \kappa} = \sqrt{1 + \beta_L -2( \lambda_x \, \cos \kappa_x
                                +\lambda_y \, \cos \kappa_y ) }  , 
\end{eqnarray}
where ${\bf \kappa} =(\kappa_x, \kappa_y)$ is the normalized wavevector.
For low-amplitude driving field, an rf SQUID exhibits a resonant magnetic
response at a particular frequency $\omega_{SQ} = \omega_0 \sqrt{ 1 +\beta_L }$,
which lies in the middle of the linear band whose boundaries are determined by the minimum 
and maximum values of Eq. (\ref{2.04}).
The SQUID resonance can be tuned either by varying the amplitude of the 
ac driving field or by varying the magnitude of a dc field that generates
the flux threading its ring. The resonance shift 
due to a dc flux bias has been actually observed in experiments with low$-T_c$, single 
rf SQUIDs in the linear regime \cite{Jung2013,Ghamsari2013}.


\section{Multistability, Permeability, Tunability}
The maximum total current of the SQUID metamaterial, defined as 
\begin{eqnarray}
\label{2.123}
   i_{max}=max_T \left\{\frac{1}{N^2} \sum_{n,m} i_{n,m} (t) \right\} ,
\end{eqnarray}
where $N^2$ is the total number of SQUIDs in the array and $T=2\pi/\Omega$,
is calculated numerically as a function of the driving frequency $\Omega$.
Such a typical characteristic curve is shown in Fig.2(a), where hysteretic effects are
clearly evident. The ragged part of the curve indicates that the SQUID metamaterial,
when the high-amplitude state becomes unstable, goes through a sequence of different 
intermediate states until the low-amplitude state is reached.
In other cases (not shown), the high-amplitude state remains stable for a considerable
frequency interval. Furthermore, the low- and high-amplitude states, that are 
simultaneously stable in the region of hysteresis, correspond to almost uniform
states in which individual SQUIDs are synchronized up to a very high degree 
\cite{Lazarides2013b}.
\begin{figure}[h!]
\centerline{\epsfig{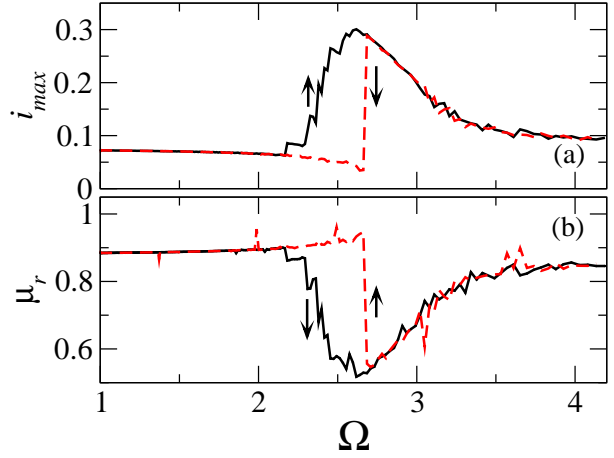}}
\caption{(a) The maximum total current $i_{max}$ as a function of the driving frequency
$\Omega$ for a SQUID metamaterial with $N=20$, $\alpha=0.002$, 
$\beta=1.27$,  $\phi_{ac} =0.1$, and $\phi_{dc} =0$.
(b) The corresponding relative magnetic permeabilities $\mu_r$ as a function of
$\Omega$. Multiple-valued magnetic response is observed in the bistability region. 
}
\label{fig:2}
\end{figure}
The magnetic response of the SQUID metamaterial in a particular state can be 
calculated in terms of the magnetization \cite{Lazarides2013b,Jung2013}. 
Then, using fundamental realations, the relative magnetic permeability $\mu_r$
for the SQUID metamaterial takes the form
\begin{equation}
\label{4.05}
  \mu_r = 1 +\kappa \frac{<i>}{<\phi_{ext}>} , 
\end{equation}
where $<i> \equiv \frac{1}{N^2} \sum_{n,m} <i_{n,m}>_T$, with the brackets $< >$ 
indicating temporal averaging (over the period $T$), and $\kappa$ is a coefficient
that depends on the SQUID and geometrical parameters.
Simultaneously stable SQUID metamaterial states respond differently to the
external field and therefore exhibit different $\mu_r$, as shown in Fig. 2(b),
where $\mu_r$ as a function of $\Omega$ is plotted from Eq. (\ref{4.05}). 
Note that $\mu_r$ may become negative at particular frequency regions, where the 
SQUID metamaterial behaves extremely diamagnetically.

The total energy of the SQUID metamaterial, in units of the Josephson energy $E_J$,
reads
\begin{eqnarray}
\label{3.01}
   E_{tot} =\sum_{n,m} \left\{ \frac{\pi}{\beta} 
       \left[ \dot{\phi}_{n,m}^2 +( \phi_{n,m} -\phi_{ext} )^2 \right] \right.
           -\cos(2\pi \phi_{n,m}) 
\nonumber \\  
      -\frac{2\pi}{\beta}  \left[ 
  \lambda_x ( \phi_{n+1,m} -\phi_{ext} )( \phi_{n,m} -\phi_{ext} ) \right.
\nonumber \\  \left. \left. 
  +\lambda_y ( \phi_{n,m} -\phi_{ext} )( \phi_{n,m+1} -\phi_{ext} ) \right] \right\}. 
\end{eqnarray}
The total energy $E_{tot}$, averaged over one period $T$ of evolution, is defined as
\begin{equation}
\label{2.07}
   <E_{tot}>_T =\frac{1}{T} \int_0^T d\tau E_{tot} (\tau) .
\end{equation}
The latter quantity is used for monitoring the tuneability of the linear flux wave band 
with a dc applied magnetic field. This quantity is then mapped on the 
dc flux bias $\phi_{dc}$ - driving frequency $f$ plane as shown in Fig. 3,
where dark (red) regions indicate regions with high energy concentration that correspond
to the linear band. It is clearly observed that the linear band shifts periodically in
$\phi_{dc}$ with period one flux quantum.
The calculations with Eqs. (\ref{2.01}) reproduce very well recent experimental results 
where the complex transmission $S_{21}$ is measured and mapped on the $\phi_{dc}$-$f$
plane \cite{Butz2013a,Trepanier2013}, where regions of reduced $S_{21}$ correspond to 
regions with high energy concentration as in Fig. 3.

SQUID metamaterials support magnetoinductive flux waves just as conventional metamaterials
do. In the linear regime, flux waves are allowed to be transmitted only at frequencies 
within the linear band, Eq. (\ref{2.04}). However, with increasing amplitude of the 
alternating field, nonlinearity becomes important, and the opening of nonlinear transmission
bands is observed. The efficiency of these bands is determined primarily by the level of 
nonlinearity which in turn is controlled by the amplitude of the applied field.
Thus, there is the possibility to switch on and off a nonlinear transmission band by
varying the field amplitude.

\vspace{-0.5cm}
\begin{figure}[h!]
\centerline{\epsfig{figure=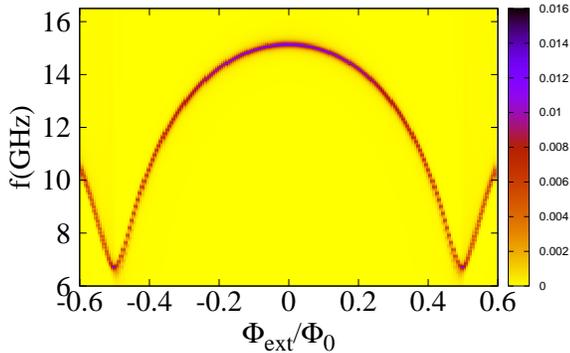,width=80mm}}
\caption{Density plot of the total energy, averaged over a period $T$ of evolution,
$<E_{tot}>_T$, as a function of the dc flux bias $\phi_{dc}=\Phi_{dc}/\Phi_0$
and the driving frequency $f$, for $N=11$, $\beta_L \simeq 0.7$,
$\gamma =0.009$, $\lambda_{x,y} =-0.01$, and $\phi_{ac}=1/5000$.
}
\label{fig:3}
\end{figure}
\section{Conclusions}
The dynamics of SQUID metamaterials exhibit interesting properties, like multistability 
and multiple magnetic responses, negative permeability and wide-band tuneability.
The presence of low-levels of disorder introduced through the critical currents seem 
to favor hysteretic effects and synchronization between individual SQUIDs.
Moreover, the tuneability patterns obtained in recent experiments are fairly well
reproduced by the model equations, which exhibit relatively small qualitative
differences from one system to another. Controllable transmission through nonlinear 
bands seems possible with the variation of the applied field amplitude.

\begin{acknowledgement}
This research was partially supported
by the European Union's Seventh Framework
Programme (FP7-REGPOT-2012-2013-1) under grant agreement n$^o$ 316165, and
by the Thales Project "MACOMSYS",
cofinanced by the European Union (European Social Fund - ESF)
and Greek National Funds through the Operational Program
"Education and Lifelong Learning" of the National Strategic Reference Framework
(NSRF)  "Investing in knowledge society through the European Social Fund".
\end{acknowledgement}

\bibliographystyle{IEEEtran}

\end{document}